\documentclass{aip-cp}

\usepackage{etoolbox,epsfig}
\makeatletter
\patchcmd{\@@tablenote}{\xdef}{\protected@xdef}{}{}
\makeatother

\usepackage[numbers]{natbib}
\usepackage{rotating}

\begin{document}

\title{How to Test the Two-Families Scenario}

\author[aff1]{Prasanta Char}
\author[aff2,aff1]{Alessandro Drago}
\author[aff2,aff1]{Giuseppe Pagliara}

\affil[aff1]{INFN Sez. Ferrara, Via Saragat 1, 44122 Ferrara, Italy}
\affil[aff2]{Dipartimento di Fisica e Scienze della Terra, Univ. Ferrara, Via Saragat 1, 44122 Ferrara, Italy}

\maketitle

\begin{abstract}
We shortly summarize the two-families scenario in which both hadronic stars and strange quark stars can exist and we describe the main predictions one can obtain from it. We then concentrate on the observables that most likely will be measured in the near future, i.e. masses, radii, tidal 
deformabilities and moments of inertia and we present a list of objects that are candidate strange quark stars in this scheme. We show that the estimates of the radii derived up to now from observations are all compatible with the two-families scenario and in particular all the objects having large radii can easily be interpreted as strange quark stars.
\end{abstract}

\section{INTRODUCTION}

The two-families scenario has been introduced in 2014 \cite{Drago:2013fsa,Drago:2014oja} in order to interpret two conflicting observational indications: the existence of massive "neutron" stars, with masses exceeding 2 $M_\odot$ \cite{Demorest:2010bx,Antoniadis:2013pzd,Cromartie:2019kug} and the possible existence of objects with very small radii, of the order of 11 km or less \cite{Ozel:2016oaf,Shaw:2018wxh}. The scenario is based on the idea that hadronic matter gets soft at large densities due to the production of resonances. In that way very compact configurations, made only of hadrons, can be obtained (first family), but their maximum mass cannot reach 2 $M_\odot$. 
Instead pure quark matter can be very stiff \cite{Alford:2006vz,Kurkela:2009gj,Weissenborn:2011qu} and therefore the most heavy compact objects in this scenario are strange quark stars and their radii are not very small (second family). Indicatively the maximum mass of hadronic star 
can be of the order of $(1.5-1.6) M_{\odot}$.
The idea of two-families of objects, one containing quarks and the other not, is not new and it was discussed e.g. in \cite{Berezhiani:2002ks,Bombaci:2004mt}. The new idea introduced in 2014 was that the small-radii branch was not the one made of quarks, but the one made of hadrons. This also implies that the radii of the most massive stars need not to be small since they are strange quark stars.

We start by summarizing the main ideas on which the two-families scenario is based, a more detailed analysis can be found for instance in \cite{Drago:2015cea,Drago:2015dea,DePietri:2019khb}.

\subsubsection{Production of resonances and softening of the Equation of State}
The first idea is that by increasing the density more and more resonances are produced and therefore hadronic matter becomes more and more soft. In \cite{Drago:2013fsa,Drago:2014oja} we did concentrate in particular on the production of $\Delta$-resonances, since quite easily they appear at densities of the order of $2-3$ times nuclear matter saturation density $\rho_0$ if the symmetry energy does not grow too rapidly with density \cite{Lattimer:2012xj}. This result has been confirmed by many other papers, see for instance \cite{Cai:2015hya,Li:2018qaw}, although up to now a microscopic non-relativistic calculation of the production of $\Delta$-resonances in that density range is still missing. The production of $\Delta$-resonances {\it{does not}} solve the hyperon-puzzle problem, because these resonances contribute to the softening of the EoS in a way similar to the production of hyperons. Instead it helps in getting small radii without the need of producing hyperons.

\subsubsection{Quark deconfinement}
The mechanism of quark deconfinement included in the two-families scenario is based on the idea of meta-stability of matter. According to Bodmer-Witten hypothesis \cite{Bodmer:1971we,Witten:1984rs} ordinary matter is meta-stable respect to the formation of strange quark matter, the reason why it does not decay immediately is that {\it{if strange quarks are not already present}} the production of strangeness via multiple and simultaneous weak reactions is exceedingly slow. The basic idea is that until in the system the density of strange quarks (inside hyperons and/or potentially inside kaons) does not reach a critical value the process of quark deconfinement will not start (there are possible deviations from this simple scheme, for instance quark deconfinement could start already with only two flavours and then continue through strangeness production till the conditions for the Bodmer-Witten hypothesis are satisfied. For simplicity we have mainly analyzed the previous scheme). After quark deconfinement starts by nucleating the first droplets of strange quark matter, the deconfinement transition becomes fast and it can be described by using the standard hydrodynamical framework of combustion \cite{Lugones:1994xg,Drago:2005yj,Horvath:2007tv,Niebergal:2010ds,Drago:2015fpa}. A very important result is that the acceleration of the combustion front, due to hydrodynamical instabilities, allows a rapid conversion to quark matter of the bulk of the star, but it stops at a density determined by the so-called Coll condition \cite{Coll:1976} and at lower densities the process is purely diffusive and much slower. This result was obtained in numerical simulations \cite{Herzog:2011sn,Pagliara:2013tza} and has been confirmed and interpreted analytically in \cite{Drago:2015fpa}.

\subsubsection{Structure of quark stars}
As explained before, QSs are allowed if the Bodmer-Witten hypothesis is correct,
i.e. if the true ground state of strongly interacting matter
is strange quark matter \cite{Bodmer:1971we,Witten:1984rs}. In turn, this implies the existence 
of a new minimum of the energy per baryon $e_0/n_0$ at a finite value of the baryon density $n_0$ and whose value is smaller than the energy per baryon of Iron ($<930$ MeV). Consequently, QSs are self-bound, at their surface the pressure vanishes but the energy density is large and, as it has been observed in \cite{Haensel:1986qb}, the adiabatic index diverges at the surface of the star.
This feature leads to very stable configurations and it allows 
for the existence of very massive QSs (of the order of $2M_{\odot}$ as it has been suggested long time ago in the seminal paper of Witten \cite{Witten:1984rs}).
Notice that the evidence of the existence of stars as massive as $2M_{\odot}$ is quite recent (\cite{Demorest:2010bx}).
In this respect, the idea that massive compact stars
could be QSs is rather old and it can be regarded as 
a theoretical prediction, see also the results from pQCD calculations \cite{Kurkela:2009gj}.
Another remarkable point about QSs is that they can reach large maximum masses even if the speed of sound is taken to be smaller/equal
than the asymptotic limit of $1/3$ which is expected to be realised at very 
high density (due to the QCD asymptotic freedom).
This is an important difference between the two-families
scenario and the scenario of twin stars which is obtained 
in the presence of a very strong first order phase transition from 
hadronic matter to quark matter \cite{Paschalidis:2017qmb,Montana:2018bkb}.
In the latter case, to reach large maximum masses,
one is forced to assume that the speed of sound is very close to the causal limit. One should then clarify which could be the physical mechanism which reduces the (squared) speed of sound to $1/3$ at very large densities.

Let us turn now to the structural properties of QSs.
For the present discussion we will use the most simple
model for the quark matter EoS i.e. a constant speed of sound
EoS with parameters $c^2_s$ and $e_0$ which correspond
respectively to the squared speed of sound (here fixed to its asymptotic limit of $1/3$) and to the energy density at vanishing pressure. The relation between pressure and 
energy density reads: $p=c^2_s(e-e_0)$. The parameter $e_0$
is unknown and it can be regarded as a free parameter provided that it is not smaller than the energy density of nuclei i.e.
$e_0 \geq 150$ MeV/fm$^3$ , see \cite{Lattimer:2010uk}.

\section{IMPLICATIONS OF QUARK DECONFINEMENT FOR EXPLOSIVE PHENOMENA}

The possibility of transforming a hadronic star into a strange quark star has many implications for explosive phenomena as SNe and GRBs.
As discussed above, the mechanism to trigger quark deconfinement is based on reaching a critical fraction of strangeness in the center of the star and the simplest way to do so is to consider mass accretion.
In the following we summarize the main astrophysical possibilities.

\subsubsection{Supernovae}
The idea that quark deconfinement can help supernovae to explode is an old one \cite{Gentile:1993ma,Drago:1997tn} and is often connected with the possibility that a mixed phase of quarks and hadrons can form during the collapse and that the collapsing center of the star rapidly crosses the soft mixed phase to bounce on the hard pure quark matter phase: this second bounce will finally trigger the explosion \cite{Sagert:2008ka}. Recently that idea has been revisited to provide a new mechanism for the explosion of very massive progenitors \cite{Fischer:2017lag}, which are likely at the origin of the most massive compact stars. The idea that quark deconfinement can help supernovae to explode is also present in the two-families scenario, but in a slightly different form: what helps the explosion is not the bounce on the stiff pure quark matter phase (since no mixed phase forms in this scenario), but the rapid combustion of the central part of the star. The process of quark deconfinement is strongly exothermic \cite{Drago:2004vu} and releases a neutrino flux having a luminosity so large 
\cite{Pagliara:2013tza} that it can in principle revitalize a failed supernova explosion. Unfortunately no explicit simulation of this mechanism has been performed so far, but it could provide a mechanism similar to the one discussed in \cite{Fischer:2017lag} to explain the formation of the most massive compact stars which, in the two-families scenario, are strange quark stars. It is interesting to notice that there is some evidence of a bimodal distribution of the masses of neutron stars \cite{Antoniadis:2016hxz}, but standard mechanisms for SN explosion do not provide a justification for that. The possibility of a special mechanism, based e.g. on quark deconfinement, and active in the case of massive progenitors could help in supporting the statistical evidence, although at the moment the latter is rather weak.

\subsubsection{Long Gamma-Ray Bursts}
One of the first suggestions of the possible relevance of quark deconfinement in astrophysical explosions was to link long Gamma Ray Bursts to the formation of deconfined quarks in a compact star \cite{Bombaci:2000cv}, by noticing that it is a strongly exothermic process. The difficulty in assessing the relevance of that idea is due to the need to find signatures of that process. A possible signature is associated with the time delay between the moment of explosion of the supernova (either successful or partially failed) and the moment of quark deconfinement. This time delay can be linked to delays in the EM signals observed in GRBs \cite{Berezhiani:2002ks,Drago:2005rc}, but other explanations are possible. 

\subsubsection{Short Gamma-Ray Bursts and mergers}
In the last years a great number of analysis have been dedicated to study the relation between mergers of two neutron stars (or of a black hole and a neutron star) and short Gamma-Ray Bursts. Within the two-families scenario three types of mergers can take place: two hadronic stars, a hadronic star with a strange quark star, two strange quark stars. The phenomenology is rather complex and it has been analyzed in \cite{Drago:2017bnf} and in \cite{DePietri:2019khb}. The main prediction is related to the mergers of two hadronic stars: since the maximum mass of a hadronic star in the two-family scenario is significantly smaller than $2 M_\odot$, the threshold mass (i.e. the mass above which a direct collapse to a BH takes place) is smaller than in the 1-family scenario and it turns out to be of the order of $2.5 M_\odot$ (while in the 1-family case it is larger than $3 M_\odot$). This implies that the model predicts the possibility of direct collapses to BH for masses smaller than the one observed in GW170817. The latter, in the two-families scenario, is interpreted as due to the merger of a hadronic star with a strange quark star. There are other implications of this scenario, related e.g. to the amount of mass emitted during the merger (relevant for the kilonova signal) and they are discussed in detail in \cite{DePietri:2019khb}.

\section{OBSERVATIONAL CONSTRAINTS FROM MASSES, RADII, TIDAL DEFORMABILITIES AND MOMENTS OF INERTIA}

Astrophysical observations can help to constrain the equation of state of dense matter inside a compact star. The discoveries of massive pulsars have already ruled out many softer hadronic equations of state which do not give a maximum TOV mass over $2 M_\odot$ within the one-family scenario. The observation of gravitational waves from the binary neutron star merger event GW$170817$ has provided the information on the tidal deformability which can be translated into a radius constraint. Latest analysis of that data provides the tidal deformability for a $1.4 M_\odot$ star, $\Lambda_{1.4}=190^{+390}_{-120}$ \cite{TheLIGOScientific:2017qsa, Abbott:2018exr,Fasano:2019zwm}. This indicates towards a softer EOS in the hadronic phase. This gives us an unique opportunity to probe the two-families scenario: in
\cite{Burgio:2018yix} is was shown that the two-families scenario (as well as the twin-stars scenario)
offer the possibility of having rather large values of $\tilde{\Lambda}$ while allowing for the existence 
of stars with radii smaller than about $11$km.

First results from NICER are awaited on the observations of thermal x-ray emissions from the polar caps of rotation-powered millisecond pulsars \cite{nicer}. NICER uses the technique called ``pulse profile modeling" which can provide simultaneous measurements of stellar compactness and radius \cite{Watts:2016uzu}. With these measurements certain regions of mass radius can be strictly constrained. Preliminary results indicate towards a larger radius for at least one of the sources.

In future, Square Kilometer Array (SKA) will also start observing the pulsars in a binary system with very precise timing \cite{Watts:2014tja}. This will help to constrain the moment of inertia of the highly relativistic pulsar PSR J$0737-3039$A within $10\%$ accuracy level. When we get a few of such measurements, we can devise a schematic framework to distinguish among the families of compact stars.  If the radius of a 1.4 M$_\odot$ star ($R_{1.4}$) (i.e. with a mass so low that it could not be a quark star) is greater than $13$ km or equivalently, by using the formualae in \cite{Bejger:2002ty}, I$_{45} \geq 1.88$, it indicates
that its central density $\rho_{max} \leq 3 \rho_0$ and therefore the star can be purely nucleonic \cite{Lonardoni:2014bwa} . For the range, $11.5$km $\leq R_{1.4} \leq 13$km or $1.55 \leq I_{45} \leq 1.88$, the central density ($\rho_{max} \leq 5 \rho_0$) is higher than for the purely nucleonic star and  
one can argue for the case of a hybrid or a hyperonic star \cite{Nandi:2017rhy,Maslov:2015msa}. Finally, when $R_{1.4} \ll 11.5$ km or $I_{45} \ll 1.55$, the two-family or twin-star interpretation is inevitable.

As we already have precise measurements of the mass of many pulsars, we can also group the possible candidates for quark stars i.e. stars having a mass larger 
than $(1.5-1.6) M_{\odot}$ listed in Table 1. 
Instead in Table 2 we present a short list of objects having large radii that in the two-families scenario would also be interpreted as
strange quark stars.

\begin{table}[h]
 \centering
 \begin{tabular}{cccc}
 \hline
  Name of the Pulsar & Frequency (Hz) & Mass (M$\odot$) & Notes \\
  \hline
  J$0348+0432$ & 25.561 & 2.01(4) &  From \cite{Ozel:2016oaf}\\
  J$0453+1559$ & 21.843 & 1.559(5) & \\
  J$0621+1002$ & 34.657 & ${1.53}^{+0.10}_{-0.20}$ & \\
  J$0740+6620$ & 346.532 & ${2.17}^{+0.11}_{-0.10}$ & Millisecond Pulsar (MSP) - massive white dwarf (WD) \\
  J$0751+1807$ & 287.458 &  1.64(15) & MSP - He WD\\
  J$1012+5307$ & 190.267 & 1.64(22) & MSP - He WD\\
  J$1614-2230$ & 317.379 & 1.908(16) & MSP - massive WD\\
  J$1903+0327$ & 465.135 & 1.667(21) &  1 $M_\odot$ companion\\
  J$1946+3417$ & 315.444 & 1.828(22) & MSP - He WD\\
  J$2222-0137$ & 30.471 & 1.76(6) & Recycled pulsar - massive WD \\
  J$1748-2446$I & 104.491 & ${1.91}^{+0.02}_{-0.10}$ & From \cite{Lattimer:2012nd}  \\
  J$1748-2446$J & 12.447 & ${1.79}^{+0.02}_{-0.10}$ &  \\
  J$0621+1002$  & 34.657 & ${1.70}^{+0.10}_{-0.17}$ & Reanalyzed to ${1.53}^{+0.10}_{-0.20}$ (Kasian 2012 PhD thesis) \\
  J$0437-4715$ & 173.688 & 1.76(20) & Reanalyzed to 1.44(7) \cite{Reardon:2015kba} \\
  J$2043+1711$ & 420.189 & 1.85(15) & Reanalyzed to ${1.38}^{+0.12}_{-0.13}$ \\ 
  4U $1702-429$ & 329 & 1.9(3)  & \cite{Nattila:2017wtj}; frequency from \cite{Patruno:2017oum} 
 \end{tabular}
\caption{Candidates for quark star in two-family scenario (see \cite{Ozel:2016oaf, Lattimer:2012nd}, and references therein)}
\end{table}

It is interesting to notice the large number of objects rotating at about 300 Hz. In \cite{Patruno:2017oum}, it was suggested that in LMXBs two sub-populations of fast spinning objects exist: one centered at about 300 Hz and with a large spread in frequency and a second more peaked and centered at about 575 Hz. Since the transition from hadronic star to strange quark star due to mass accretion in e.g. LMXBs increases the moment of inertia \cite{Pili:2016hqo}, conservation of angular momentum during the transition can shift the rotational frequency of the object to a lower value. This argument could possibly explain why so many candidate strange quark stars rotate at about 300 Hz. 

We can infer on the nature of quark star matter from these observations. Formation of Cooper pairs by the BCS mechanism is expected in the deconfined matter. But, if it contains only the color-flavor-locked (CFL) phase which has very low both shear and bulk viscosity, the rotating star would be unstable against the r-modes. Thus, it can be two flavor color superconducting (2SC), or otherwise just unpaired quark matter. There is also a possibility of crystalline color superconducting phase on the outer layer encapsulating the CFL core of the star \cite{Mannarelli:2014ija}.
Unfortunately there is no calculation of the viscosity for that type of quark matter.

\begin{table}[]
    \centering
    \begin{tabular}{ccc}
       Name  & Radius (km) & Notes \\
       \hline
       RX J$1856.5-3754$  & ${12.1}^{+1.3}_{-1.6}$ & \cite{Potekhin:2014hja}  \\
       J$0437 - 4715$ & ${13.1}^{+0.9}_{-0.7}$ & \cite{Gonzalez-Caniulef:2019wzi} \\
        4U $1702-429$ & ${12.4}^{+0.4}_{-0.4}$  & \cite{Nattila:2017wtj}
    \end{tabular}
    \caption{Stars with larger radii}
    \label{tab:my_label}
\end{table}

\subsubsection{Implications for the quark and for the hadronic equations of state}
We can fix $e_0$ by requiring that the maximum mass of QSs is $\gtrsim 2.3 M_{\odot}$
as suggested by the analyses of sGRB featuring an extended emission \cite{Lasky:2013yaa,Lu:2015rta}. In Fig.~1 we display the result for the $M-R$ relation (left panel) and the $M-\Lambda$ relation (right panel) of QSs and for hadronic stars (equation of state taken from \cite{Drago:2014oja}). For comparison, we also show some of the most recent observational limits. Concerning radii,
we have included the most recent analysis on PSR J0437-4715 \cite{Gonzalez-Caniulef:2019wzi}, 
on 4U 1702-429 \cite{Nattila:2017wtj} and on low mass X-ray binaries in quiescence (QLMXBs)\cite{dEtivaux:2019cnf}. On the other hand, the only available limit on $\Lambda$
comes from GW170817 \cite{Abbott:2018exr,Fasano:2019zwm} which, in turn,  translates also into a limit on the radius \cite{Abbott:2018exr}. In particular, we have indicated in the figure
the limits for the most massive of the two stars in the binary, which, within our scheme, is 
indeed interpreted as a QS \cite{Drago:2017bnf,Burgio:2018yix,DePietri:2019khb}.

\vskip 0.3cm

\begin{figure}[!ht]
	\begin{centering}
		\epsfig{file=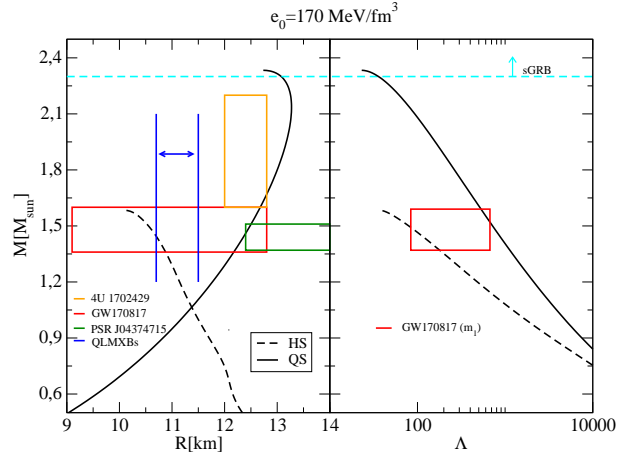,height=8cm,width=6cm,angle=-90}
		\caption{Right panel: mass-radius curve of QSs and HSs and observational constraints 
		on the maximum mass of compact stars (as obtained by the analysis on sGRBs) and on the radii (as obtained from the direct modelling of the X-ray spectra of 
		PSR J0437–4715 ($1\sigma$ level), of 4U 1702−429 ($1\sigma$ level) and of different LMXBs in quiescence ($1\sigma$ level)). We also display the constraints on the radius of the most massive star in the merger event GW170817 ($90\%$ confidence interval). Right panel: mass - tidal deformability relation of QSs and observational constraint obtained from GW170817 ($90\%$ confidence interval).   }
	\end{centering}
\end{figure}

It is remarkable that our curves for QSs, which depend only on $e_0$ within the simple constant speed of sound EoS, are consistent
with all the available observational constraints suggesting large radii. Instead 
the constraints on radii obtained from LMXBs in quiescence \cite{dEtivaux:2019cnf} suggest a central value of radius of $11.1$ km. Such a small value of the radius 
can be explained only in two ways: the two-families scenario in which such 
small stellar objects are interpreted as hadronic stars \cite{Drago:2013fsa,Drago:2015cea,Burgio:2018yix} or the twin-stars scenario 
in which they would be interpreted as hybrid stars (with a strong first order phase transition from hadronic matter to quark matter) \cite{Paschalidis:2017qmb}.

Concerning the observation of  gravitational waves (right panel),
notice that the limit on $\Lambda$ constrains the mass of the QS to be larger than about to $1.52M_{\odot}$ corresponding to a mass ratio $q$ smaller than about $0.8$. Therefore, in the two families scenario, GW170817 
has been produced by a rather asymmetric system.
This could help explaining the large amount of material ejected by the merger and powering the kilonova.

\section{CONCLUSIONS}
We have reviewed the two-families scenario and its phenomenological consequences relatated to the coexistence of hadronic stars and quark stars.
The equation of state of hadronic matter and of quark matter are rather uncertain
and therefore it is difficult to provide exact theoretical predictions e.g. on the maximum mass of quark stars or on the $R_{1.4}$ for hadronic stars. 
However, if the tension between massive stars and small radii will be confirmed 
by future more precise measurements, the one-family scenario will be disfavoured
whereas
the two-families scenario will represent, together with the twin-star scenario, a viable solution. In both cases the presence of quark matter is a key ingredient but the two scenarios make completely different predictions for stars with large masses (close to $2.3M_{\odot}$) : while in the two-families scenario those stars are quark stars
and have radii necessarily larger than $R_{1.4}$, in the twin-star scenario those stars are hybrid stars and their radii is significantly smaller than $R_{1.4}$ \cite{Paschalidis:2017qmb}. In turn, this translates into very different values of the tidal deformabilities (which could affect the dynamics of the merger)
and very different values of the moment of inertia.

\nocite{*}
\bibliographystyle{aipnum-cp}%
\bibliography{readme}%

\end{document}